# Quantum rotator: some examples in modern spirit

Yuri Kornyushin

*Maître Jean Brunschvig Research Unit, Chalet Shalva, Randogne, CH-3975, Switzerland*
jacqie@bluewin.ch

Rotation of the nucleus and rotation of the electronic cloud of the atom/ion were considered. It was shown that these rotations are not practically possible. Rotation of the cloud of delocalized electrons and ionic core of a fullerene molecule and these of the ring of a nanotube were discussed. It was shown that the rotation of the cloud of delocalized electrons of a fullerene molecule is possible and it goes in a quantum way when temperature is essentially lower than 40 K. Rotation of the ion core of a fullerene molecule is possible in a classical way only. The same should be said about rotations in the ring of a nanotube.

**1. Introduction**

Rotating body is a subject of quantum mechanics [1]. We regard here rotations around principal axes of a body [1 – 3]. In general case there are 3 principal moments of inertia of a body [1 – 3]. Quantum levels of rotator are [1]

$$E_J = (\hbar^2/2I)J(J + 1). \qquad (1)$$

Here $\hbar$ is Planck constant divided by $2\pi$, principal moment of inertia of a body [2] $I = \int\rho(\mathbf{r})x^2 dV$ (integral is taken over the body volume $V$, $\rho(\mathbf{r})$ is the mass density, and $x$ is the perpendicular distance to the principal axis of rotation); $J$ is zero or integer.

When the mass density is constant along the body, one can see that the moment of inertia of a body is $I = m\langle x^2\rangle$ (here $m$ is the mass of a body, $\langle\ldots\rangle$ denotes averaging over the volume of a body).

Now let us consider a body of a shape of a spherical layer, $R_i \leq r \leq R$. For this spherically symmetric object there is only one principal moment of inertia. Calculating it, $I = \int\rho(\mathbf{r})x^2 dV$, one gets

$$I = 0.4mR^2 + 0.4mR_i^3[(R + R_i)/(R^2 + RR_i + R_i^2)]. \qquad (2)$$

We shall consider further on several rotating objects.

**2. Atoms and ions**

Atom (ion) consists of a nucleus and an electronic cloud. We shall consider a nucleus in the nuclear liquid drop model [4]. We assume that the density of a mass in the nucleus is constant in the sphere of a radius $R$. It follows from Eq. (2) that, as $R_i = 0$ for the nucleus, $I = 0.4mR^2$ and

$$E_J = 1.25(\hbar^2/mR^2)J(J + 1) = 1.25(\hbar^2/mr_0^2A^{2/3})J(J + 1). \qquad (3)$$

For the nucleus of a carbon atom $m = 2\times10^{-26}$ kg [5]. The radius of a nucleus $R = r_0A^{1/3}$ [1] (here $A$ is the number of the nucleons in the atom, $r_0 = 1.2\times10^{-15}$ m). For carbon $A = 12$ and according to Eq. (3) $E_J = 0.5754J(J + 1)$ MeV. This is an enormous value, inaccessible. No rotation is practically possible.

Let us consider electron cloud of the atom/ion in the nuclear liquid drop model also [6]. In this model the radius of the electronic cloud of the atom/ion is as follows [6]

$$R_e = 7.37[N^{2/3}/(5Z - 2N)](g\hbar^2/m_e e^2). \quad (4)$$

Here $N$ is the number of the electrons in the cloud, $Z$ is the number of the protons in the nucleus, $g = 0.317$ [6], and $m_e$ is the electron mass. For the atom/ion Eqs. (1), (2), and (4) yield:

$$E_J = 1.25(\hbar^2/Nm_e R_e^2)J(J + 1) = 0.0230131[(5Z - 2N)^2/g^2 N^{7/3}](m_e e^4/\hbar^2)J(J + 1). \quad (5)$$

For a neutral atom of carbon $N = Z = 6$ and we have $E_J = 6.8056(m_e e^4/\hbar^2)J(J + 1)$. It is worthwhile to note that $m_e e^4/\hbar^2 = 27.21$ eV [1]. The value of $E_J$ for the electronic cloud is also of a very large number. So rotation of this object is hardly possible.

### 3. Fullerene molecule

For the delocalized electron cloud of a fullerene molecule ($C_{60}$) $N = 240$, $R_i = 0.279$ nm, and $R = 0.429$ nm [7]. For these numbers Eq. (2) yields $I = 21.5364 m_e$ (in $m_e$nm$^2$ units). For these numbers Eq. (1) yields $E_J = 0.001771J(J + 1)$ eV. For $J = 1$ we have $E_J = 0.003542$ eV. This value corresponds to 41.1 K. When temperature is essentially smaller than 40 K the behavior of this rotator has a quantum character. Anyway rotation of the electronic cloud of a fullerene molecule is fairly possible.

The ionic core of a fullerene molecule has radius $R_f = 0.354$ nm [7]. Taking into account that as was mentioned before the mass of a carbon atom is $2\times 10^{-26}$ kg, and that there are 60 carbon atoms in a fullerene molecule, we have $E_J = 2.31\times 10^{-7}J(J + 1)$ eV. For $J = 1$ we have $E_J = 4.62\times 10^{-7}$ eV. This corresponds to 0.005362 K. This quantity is so small that quantum behavior is utterly impossible. This object rotates always classically.

### 4. Ring of a nanotube

Let us consider a ring of a nanotube of a radius $R$. This ring has a length $2\pi R$. We assume that the radius of a ring $R$ is essentially larger than the diameter of a nanotube $d$ ($d$ is typically about 1.4 nm [8]). We accept also as in [8] that the number of the delocalized electrons in a nanotube is $N_n = 670$ nm$^{-1}$. Then Eq. (1) yields

$$E_J = (\hbar^2/4\pi m_e N_n R^3)J(J + 1). \quad (1)$$

For $R = 100$ nm we have $E_J = 2.846\times 10^{-11}$ eV. This is an extremely small quantity. This means that rotation of this object is fairly classical. The same should be said about rotation of the ion core of the regarded ring, and rotation of the ring as a whole around the axis, going through the middle of the ring.

**References**

1. Landau, L. D., and Lifshitz, E. M., 1987, *Quantum Mechanics* (Oxford: Pergamon).
2. Goldstein, H., 1950, *Classical Mechanics*, (Cambridge Mass.: Addison-Wesley).
3. Sylvester, J. J., 1852, *Phil. Mag.*, **IV**, 138.
4. Bohr, A., and Mottelson, B. R., 1975, *Nuclear Structure*, Vol. 2 (London: Benjamin).

5. 1988 CRC Handbook of Chemistry and Physics, ed. R. C. Weast (Boca Raton, FL: CRC).
6. Amusia, M. Ya., and Kornyushin, Y. 2000, *Contemporary Physics*, **41**, 219.
7. Rüdel, A., Hentges, R., Becker, U., Chakraborty, H. S., Madjet, M. E., and Rost, J. M., 2002, *Phys. Rev. Lett.*, **89**, 125503.
8. Kornyushin, Y., 2008, *Low Temperature Physics*, **34**, 838.